\documentclass[12pt]{iopart}

\usepackage[english]{babel}
\usepackage{graphicx}
\usepackage{subfigure}
\usepackage{bm}
\usepackage{iopams}
 \expandafter\let\csname equation*\endcsname\relax
  \expandafter\let\csname endequation*\endcsname\relax
\usepackage{amsmath,amssymb,amsfonts}

\usepackage{verbatim}

\begin{document}

\title[Anomalous Raman scattering in LaMnO$_3$]{Anomalous multi-order Raman scattering in LaMnO$_3$: a signature of a quantum lattice effect in a Jahn-Teller crystal}

\author{N N Kovaleva$^{1,2}$, O E Kusmartseva$^3$, K I Kugel$^{3,4}$, A~A~Maksimov$^5$, D Nuzhnyy$^2$, A M Balbashov$^6$, E~I~Demikhov$^1$, A Dejneka$^2$, V A Trepakov$^{7,2}$, F V Kusmartsev$^3$ and A M Stoneham$^8$\footnote{Deceased.}}

\address{$^1$ Lebedev Institute of Physics, Russian Academy of Sciences, Moscow, Leninsky prosp. 53, 119991, Russia}
\address{$^2$ Institute of Physics, Academy of Sciences of the Czech Republic, Prague, Na Slovance 2, 18221, Czech Republic}
\address{$^3$ Department of Physics, Loughborough University, Loughborough, LE11 3TU, UK}
\address{$^4$ Institute for Theoretical and Applied Electrodynamics, Russian Academy of Sciences, Moscow, Izhorskaya street 13/19, 125412, Russia}
\address{$^5$ Institute of Solid State Physics, Russian Academy of Sciences, Chernogolovka, Academician Ossipyan street 2, 142432, Russia}
\address{$^6$ Moscow Power Engineering Institute, Moscow, Krasnokazarmennaya street 14, 105835, Russia}
\address{$^7$ Ioffe Physical-Technical Institute, Russian Academy of Sciences, St. Petersburg, Politekhnicheskaya street 26, 194021, Russia}
\address{$^8$ London Centre for Nanotechnology, University College London, London, 17-19 Gordon Street, WC1H OAH, UK}

\ead{nkovaleva@sci.lebedev.ru}

\date{\today}

\begin{abstract}
The multi-order Raman scattering   is studied  up to a fourth order for a detwinned LaMnO$_3$ crystal. Based on a comprehensive data analysis of the polarisation-dependent Raman spectra, we show that the anomalous features in the multi-order scattering could be the sidebands on the low-energy mode at about 25 cm$^{-1}$. We suggest that this low-energy mode stems from the tunneling transition between the potential energy minima arising near the Jahn-Teller Mn$^{3+}$ ion due to the lattice anharmonicity, and the multi-order scattering is activated by this low-energy electronic motion. The sidebands are dominated by the oxygen contribution to the phonon density-of-states, however, there is an admixture of an additional component, which may arise from coupling between the low-energy electronic motion and the vibrational modes. 
\end{abstract}




\submitto{\JPCM}

\maketitle


\section{Introduction}

Anomalous Raman scattering, characterised by surprisingly strong features in the multi-order scattering, and with intensities comparable to the one-phonon scattering, seems to be a general property of rare-earth ($R$) manganites $R$MnO$_3$ \cite{Martin-Carron,Jandl,Choi,Choi1,Laverdiere}.  This phenomenon is directly related to a problem of  the existence of a new type of elementary excitation -- an orbital wave excitation -- or a new particle called ``orbiton'' in the correlated electron systems with broken symmetry \cite{SarfattStoneham,Allen_nature,Saitoh}. Usually, an orbital excitation is associated with an energy gap, induced in the two-fold degenerate electronic $e_g$ levels of  Mn$^{3+}$ ions by electron superexchange (SE) interaction \cite{Goodenough,Kanamori,KugelKhomskii,Oles} and/or Jahn-Teller (JT) electron-lattice interaction \cite{Kaplan}. Unfortunately, an uncertainty in parameters of superexchange interaction, particularly in the Mott--Hubbard value $U$ from one side, and in the JT energy $\Delta$ from the other side, is still a challenge for theoreticians and  it is not yet clear where the orbitons should be looked for in measurements. In a limit of weak JT coupling, the orbital excitations in LaMnO$_3$ are predicted to appear around 1600 cm$^{-1}$ (0.2 eV), as a result of the superexchange interaction \cite{Maekawa,Okamoto}, which indicates on the feature, observed in the second-order Raman scattering experiment in the 900--1400 cm$^{-1}$ (0.11--0.17 eV) range \cite{Saitoh}. If the contributions from SE and JT effects are considered on equal footing, the elementary excitations of the system,   according to the theory predictions, are mixed modes with both phonon and orbital character, and the phonon-orbital excitations appear as Franck--Condon sidebands at energy intervals of the Jahn--Teller vibrational energy \cite{Brink}.
Alternatively, in a limit of strong JT coupling, and strong electronic correlations at $U\rightarrow\infty$, the orbitons in LaMnO$_3$ could be self-trapped excitons with energy  $\Delta$; then the anomalous Raman scattering could be interpreted as a result of multiphonon scattering, induced by the resonant optical excitation of 2 eV across the large  JT gap, and activated by the Franck-Condon process \cite{Allen,Allen1}. However, in our comprehensive experimental (ellipsometry) and theoretical studies of LaMnO$_3$, we argue that the optical excitation at 2 eV is an intersite $d-d$ transition, governed by superexchage interaction, and evaluate the effective Mott-Hubbard parameter $U$=3.1$\pm$0.2 eV \cite{kovaleva_jetp,KovalevaLMO_prl,KovalevaLMO_prb}. On the other side, it is argued by Iliev {\it et al.} that the oxygen isotope substitution effect in the rare-earth manganites $R$MnO$_3$ provides strong evidence that the multi-order Raman spectra are of pure phonon character, rather than of orbiton or mixed orbiton-phonon character, and that the multi-order scattering has predominantly the oxygen phonon density-of-states (PDOS) origin \cite{IlievPDOS,IlievPDOS1}. These interpretations of the anomalous Raman scattering in rare-earth manganites $R$MnO$_3$ are ambiguous and are a subject of intense debate, questioning the roles of the JT electron-lattice and/or SE interactions. Moreover, even the existence of orbital excitations in these correlated systems with broken symmetry is still unproven experimentally or theoretically.

As it was shown in the earlier study by Sarfatt and Stoneham \cite{SarfattStoneham}, and in exact agreement with the prediction of the Goldstone theorem \cite{Goldstone}, octahedral complexes of Cu$^{2+}$ (and Mn$^{3+}$) ions represent striking illustration of the system with broken symmetry, where   an excitation branch should exist, which frequency tends to  zero in the limit of infinite wavelengths. The nature of the Goldstone mode in octahedral complexes of Cu$^{2+}$ (and Mn$^{3+}$) ions, in the way noted by Sarfatt-Stoneham \cite{SarfattStoneham}, is associated with the rotation by angle $\phi$ going round the rim at the minimum of a double-valued ``Mexican hat'' energy surface \cite{Stoneham}. This rotation is characteristic of the pure ``Mexican hat'' limit, and corresponds to oscillations in the configuration  of octahedral complexes, determined by the  partial occupation of the electronic  $e_g$ orbitals. This\ should result in acoustic orbital wave propagation from relative motions of different ions, where close ions have similar phases $\phi$. When the electron-lattice coupling can be considered infinite, additional terms in the Lagrangian, such as anharmonicity, could destroy the broken symmetry and the the zero-point motion round the rim at the minimum of the "Mexican hat". In such a case, the ground state is often determined by several energetically equivalent wells, and in the static case the system is confined to one of them. However, if the coupling is finite, fluctuations between the energetically equivalent wells of the distorted geometry lead to the tunneling states \cite{Stoneham,HayesStoneham,Bersuker}.

Let us consider the model materials, LaMnO$_3$ and KCuF$_3$, for a cooperative JT effect, which also exhibit antiferromagnetic ordering (at $T_{\rm N}$ of 140 K and 39 K, respectively) due to Heisenberg-type exchange interaction \cite{KugelKhomskii,FeinerOles,Oles}. In a recent optical study of KCuF$_3$ with orbital ordering temperature of $T_{OO}=800$ K, a new temperature scale $T_S$, characterised by the emergence of sharp absorption features due to the on-site $d-d$ transitions, followed by a sideband of magnon excitations, correlating with changes in the orbital ordering \cite{Paolasini} prior to the magnetic ordering, has been established \cite{Deisenhofer}. The sideband excitations were ascribed to the zone-boundary magnons (or their combinations), which became optically-active via exchange-induced dipole mechanism. It was suggested that a dynamic distortion of the environment of the Cu$^{2+}$ ions becomes static below $T_S\approx50$ K, which leads to a symmetry change, affects the orbital ordering, and paves the way for the antiferromagnetic ordering. Simultaneously, the phonon modes, involving the displacement of F$^-$ ions, soften with decreasing temperature \cite{Abbamonte}.

Similarly, LaMnO$_3$, with orbital ordering temperature of $T_{OO}=780$ K, exhibits changes in the orbital ordering-superlattice reflection, as measured by X-ray photoemission scattering in the vicinity of $T_{\rm N}$ \cite{Murakami}, and softening of the Raman-active phonon modes \cite{Granado,Granado1}. Therefore, we suggest that the interplay of superexchange and JT electron-lattice interaction in the Heisenberg antiferromagnets LaMnO$_3$  and KCuF$_3$ is accompanied by the observed anomalous sideband excitations (zone-boundary phonons or magnons), which become optically-active due to the distortion of the octahedral environment of the Cu$^{2+}$ and Mn$^{3+}$ ions. In such a case, a significant change of the orbital-ordering-superlattice structure could be indicative of the dynamic-to-static JT transition in these compounds.

In this paper, the anomalous multi-order Raman scattering is investigated for a detwinned LaMnO$_3$ crystal up to a fourth order. Here we present the results of a comprehensive study of the temperature and polarisation dependences of the first- and higher-order Raman scattering. Based on the analysis of the obtained data, we suggest that the anomalous Raman scattering is due to sidebands on the low-energy mode at about 25  cm$^{-1}$. This behaviour is anisotropic, as within our experimental accuracy we have no indication on the presence of the low-energy mode  in the $z$-polarised Raman spectra. We argue that this low-energy mode originates from the electron tunneling between the potential energy minima arising near the JT Mn$^{3+}$ ion due to the lattice anharmonicity, and the multi-order scattering is activated by this low-energy electronic motion.  Indeed, the frequency of the mode at about 25 cm$^{-1}$ is consistent with the tunneling frequencies in the dynamic limit of the JT effect in, for example, a $^2$D system of 3 $d^9$ MgO : Cu$^{2+}$ and 3 $d^1$ CaF$_2$ : Sc$^{2+}$ \cite{HayesStoneham}. In addition, there is an experimental evidence of the low-energy mode observed at around 25 cm$^{-1}$ in a far-infrared experiment\   in doped rare-earth manganites \cite{Kida,Zhukova}. The sidebands are dominated by the phonon density-of-states (PDOS) of oxygen vibrations \cite{IlievPDOS,IlievPDOS1}. Surprisingly, as we have discovered during our study, aside from the oxygen-phonon density-of-states feature, there is an additional component to the multi-order scattering, which exhibits a different behaviour. We consider that this behaviour may arise from an admixture of the electronic component in the case of a weak vibronic interaction, when the barriers between the minima are less than the characteristic vibrational frequency, and a special coupling between low-frequency electronic motions and vibrational modes takes place \cite{Bersuker,Moffitt}. We suggest that the superexchange interaction  may be crucial in driving the system of $e_g$ electrons to the regime of the dynamic JT effect, with affordable barriers for the tunneling between the potential energy minima.

\section{Experimental approach and results \label{sec:level2}}

Single crystals of LaMnO$_3$ were grown using the crucible-free
floating zone method in a mirror furnace, equipped with an
arc lamp \cite{Balbashov}. As-grown LaMnO$_3$ crystals are single phased and, below the orbital ordering temperature $T_{OO}\simeq780$ K, exhibit heavily twinned domain patterns in the orthorhombic {\it Pbnm} structure. The particular pattern and the size of the domains depend on local temperature gradients and mechanical stresses, experienced during the growth stage. We were able to remove the twins from an essential part of the sample volume following the procedure, described in detail in our previous study \cite{KovalevaLMO_prb}. As a result, the percentage of detwinning detected at some areas of the crystal surface was as high as 95 \% and over the entire surface it was at least 80 \%. The orthorhombic [010] direction was identified as perpendicular to the crystal surface by single-crystal X-ray diffraction. The sample was further characterised by magnetometry, using a superconducting quantum interference device. We determined the antiferromagnetic transition temperature $T_{\rm N}$ at 139.6 K, which is characteristic of a nearly oxygen-stoichiometric LaMnO$_3$ crystal.

The sample surface was polished to optical grade for further optical measurements. Raman spectra were obtained using a Raman-microscope spectrometer LabRAM HR (Horiba Jobin Yvon), equipped with a grating monochromator and a liquid-nitrogen cooled CCD detector. The spectrometer has high spectral resolution (0.3 cm$^{-1}$/pixel at 633 nm), large spectral range of Raman shift from 100 to 4000  cm$^{-1}$ and unique edge filter technology. HeNe laser source operating at 632.8 nm wavelength (1.96\,eV), with appropriate filters, was used for excitation under different irradiation power. The sample was mounted on a cold finger of the micro-cryostat, cooled with liquid nitrogen. We employed the near-normal back-scattering geometry with an adjustable confocal pinhole; the propagation direction of the incident and scattered light was parallel to the principal [010] orthorhombic axis of the detwinned LaMnO$_3$ crystal. The distinctive principle polarisations were recorded in the described here Raman experiment, using generic linear polarisation of the laser radiation while rotating the sample in the $xz$ plane.

Crystal symmetry determines the number of vibrational Raman-active normal modes. A factor-group analysis of the orthorhombic crystal structure of LaMnO$_3$ (space group $Pbnm$, $D^{16}_{2h}$) with 4 f.u./unit cell yields a total number of 24 (7$A_g+7B_{1g}+5B_{2g}+5B_{3g}$) Raman-active phonon modes. The Raman scattering tensor determines whether the normal modes actually
appear, and could be observed under irradiation with a polarised light at a given scattering geometry. On the oriented sample surface, the $A_g$ symmetry phonons appear in the parallel $(xx)$, $(yy)$, and $(zz)$ polarisations of the incident and scattered light, whereas the $B_{1g}$, $B_{2g}$, and $B_{3g}$ phonons appear in the crossed ($xz,zx$), ($yz,zy$), and ($xy,yx$) polarisations, respectively. Figure 1(a,c) shows the low-temperature (80 K) and room-temperature (300 K) micro-Raman spectra measured on the detwinned LaMnO$_3$  crystal for the two selected polarisations of the incident HeNe laser irradiation, along the main $z$ crystallographic direction and along the $x'$ direction, rotated by 45$^\circ$. With the incident light, polarised along the $x'$ direction, we observed $A_g+B_{1g}$ symmetry modes, whereas with the incident light, polarised along the $z$ axis, only $A_g$ symmetry modes became pronounced. At low temperature (80 K), our Raman scattering measurements allowed us to identify  six modes of $A_g$ symmetry at 148, 210, 267, 295, 452, and 495 cm$^{-1}$ and four modes of $B_{1g}$ symmetry at 197, 313, 436, and 606 cm$^{-1}$. Two strong high-frequency  modes $A_g$ (495 cm$^{-1}$) and $B_{1g}$ (606 cm$^{-1}$) correspond to the  in-plane vibrations, associated with JT in-phase anti-symmetric and symmetric stretching of the corner-shared oxygen octahedra in the LaMnO$_3$ structure, respectively \cite{Smirnova,Iliev,KovalevaYTO}. Superimposed low-temperature Raman spectra, measured for three selected polarisations in the $xz$ plane, which were normalised to the strong $A_g$ (495 cm$^{-1}$) mode, are shown in Fig. 1(b). The highest-frequency $B_{1g}$ (606 cm$^{-1}$) mode is noticeably less pronounced than the $A_g$ mode in the $x$ polarisation, whereas it is almost completely suppressed in the $z$ polarisation. Our Raman phonon spectra are similar to those of an earlier Raman study of the detwinned LaMnO$_3$ single crystal by Saitoh {\it et al.} \cite{Saitoh}.

In order to determine frequencies, widths, and integrated intensities of the phonon bands, we fitted our polarised Raman spectra, measured at temperatures 80 K $\leq T\leq 300$ K, with a set of Lorentzians. Figure 2(a,b) shows temperature dependences of the frequencies and full widths at half maximum (FWHM) of the two strong high-frequency modes, $A_g$ and $B_{1g}$. In agreement with the earlier observation by Granado {\it et al.}, the $B_{1g}$ mode shows softening below the N\'eel temperature, which was interpreted as a consequence of magnetostriction effects \cite{Granado,Granado1}. The temperature dependence was less pronounced for the frequency of the $A_g$ mode, which reveals weak polarisation dependence. At the same time, the widths of these two modes showed no distinct anomaly near the N\'eel temperature,  increasing linearly with rising temperature in the studied temperature range. The temperature coefficient of the FWHM of the anti-symmetric JT $A_g$ mode (averaged over the measured polarisations as shown in Fig. 2(b)) is about 0.07 $\pm$ 0.01 cm$^{-1}$/K, whereas for the symmetric JT $B_{1g}$ mode it is approximately half that, about 0.04 $\pm$ 0.01 cm$^{-1}$/K. Figure 2(c) shows temperature and polarisation dependences of the scaled integrated intensities of the $A_g$ and $B_{1g}$ modes. Remarkably, our Raman study on the detwinned single crystal allowed us to detect the pronounced opposite trends in the temperature behaviour of the intensities of the $A_g$ mode, measured with the laser light polarised along the $x'$ and $z$ directions (however, we noticed different behaviour in the $x$ polarisation). Moreover, the intensities of the $A_g$ and $B_{1g}$ modes show a clear kink at the N\'eel temperature at around 140 K. As it was established in the earlier study by Kr\"uger {\it et al.} \cite{Krueger}, the Raman cross-section of the one-phonon $A_g$ and $B_{1g}$ modes shows resonant behaviour  under a HeNe (1.96 eV) light irradiation, which corresponds to a maximum of the optical gap excitation  in LaMnO$_3$ at around 2 eV \cite{Tobe,Quijada,KovalevaLMO_prb}. We note that the observed anisotropy and temperature dependences of the scaled Raman intensities, measured with the incident laser light polarised along the $x'$ and $z$ directions of the detwinned LaMnO$_3$ crystal, followed the same trend as the anisotropic optical spectral weight of the 2 eV optical band, influenced by the spin-spin correlations between the Mn$^{3+}$ spins via the superexchange interaction \cite{KovalevaLMO_prb}. 

In the first-order Raman phonon scattering at higher frequencies, an additional broad band located at around $650$ cm $^{-1}$, which cannot be assigned to a normal symmetry mode, was present (see Fig. 1(a--c)). A weak band at the same frequency is distinguishable in the infrared spectra \cite{Grueniger,note}, which indicates that this excitation could be also weakly infrared-active due to the parity-breaking effect. This band is present in the Raman spectra of all isovalent Mn$^{3+}$ rare-earth $R$MnO$_3$ ($R$=La, Pr, Nd, Sm, Eu, Gd, Tb, Dy, Ho,Y) oxides around the same frequency \cite{Laverdiere,IlievPDOS1}, demonstrating its general nature. As follows from our polarisation spectra, displayed in Fig. 1(a-c), the excitation at 650 cm$^{-1}$ is not a single mode; it is composed of $A_g+B_{1g}$ symmetry modes. These could be the zone-boundary phonons, activated by disorder, which inevitably exists in these materials due to the presence of cation vacancies \cite{Tofield,Roosmalen}. Iliev {\it et al.} associates the Raman feature at 650 cm$^{-1}$ with the maximum in the phonon density-of-states (PDOS) of oxygen vibrations, accordingly to  polarisation properties and predictions, based on lattice-dynamics calculations \cite{IlievPDOS,IlievPDOS1}. However, further we show that the origin of this band may be more complicated.

The close-up of the first-order peak at 650 cm$^{-1}$ and the Raman spectra in the second order, separated from the one-phonon lattice-symmetry Raman modes and the background, is shown in the inset of Fig. 1(b). An extended structure, represented by the three broad features, is developed in the second order, which reveals strong polarisation dependence, so that only one broad feature of $A_g$ symmetry is pronounced in the $z$-polarised spectra. In the earlier Raman study by Saitoh {\it et al.} of a detwinned LaMnO$_3$ crystal, these three broad features observed at around 1008, 1170, and 1290 cm$^{-1}$ were considered as a result of the dispersion of an independent orbital excitation, induced by the superexchange interaction \cite{Saitoh}. Here we investigated in detail temperature and polarisation dependencies of the anomalous first- and second-order Raman scattering and found them dependent due to a common origin. Figure 3(a) shows the frequencies of the first-order peak at around 650 cm$^{-1}$ multiplied by two and the second-order peak at around 1290 cm$^{-1}$. A nearly one-to-one correspondence between these two frequencies in the temperature-dependent spectra of $A_g$ symmetry in the $z$ polarisation was observed. However, apparently they do not coincide in the $x$ and $x'$ polarisations as the frequencies of the first-order peak multiplied by two appear to be shifted by approximately 25 cm$^{-1}$ higher with respect to the frequencies of the second-order peak (see also the inset to Fig. 1(b), where the first- and second-order spectra are clearly delineated). At the same time, there is a clear correlation in the temperature dependencies of their peak positions, although we show a different trend in the temperature dependence in the  $x$ polarisation, which seems to demonstrate a softening-like behaviour, analogous to that of the $B_{1g}$ (606 cm$^{-1}$) mode below the N\'eel temperature (see Fig. 2(a)). Additionally, the widths of the first- and second-order peaks fall in the same range within the experimental error bars in the $x'$ and $z$ polarisations (see Fig. 3(b)). The estimated coefficient in a linear approximation  of the FWHM temperature dependence is about 0.06 $\pm$ 0.01 cm$^{-1}$/K in the studied temperature range, comparable to that of the first-order  JT $A_g$ (495 cm$^{-1}$) and $B_{1g}$ (606 cm$^{-1}$) modes. This is indicative that the anomalous scattering observed in the $x'$ and $z$ polarisations has essentially phononic character and  could be associated with the maximum in the phonon density-of-states (PDOS) of oxygen vibrations \cite{IlievPDOS}. Surprisingly, the temperature dependence of the FWHM of the first-order peak at 650 cm$^{-1}$, and its second-order replica at 1290 cm$^{-1}$, show very different behaviour in the $x$ polarisation, where the temperature coefficient of the linear temperature dependence is about 0.18 $\pm$ 0.02 cm$^{-1}$/K. Figure 3(c) shows temperature and polarisation dependencies of the scaled integrated intensities of the first-order peak at 650 cm$^{-1}$ and the second-order peak at 1290 cm$^{-1}$, which exhibit the same trends as that of the $A_g$ and B$_{1g}$ phonons (see Fig. 2(c)).

And finally, Fig. 4 shows the Raman spectra measured in the multiphonon spectral range   on our detwinned LaMnO$_3$ crystal in the $x'$ polarisation at $T$= 80 K. The spectra in the multiphonon range were observed on a background of a broad Lorentzian band (located at around 870 cm $^{-1}$, with FWHM of about 1300 cm$^{-1}$). The spectra were fitted with a set of Lorentzians, accounting for the clearly resolved intensity peaks, and the result of the fit is shown in Fig. 4. The inset shows the result of the fitting of the Raman spectra in the multiphonon spectral range,   subtracted from the contributions of the first-order lattice Raman phonons and the background. According to this analysis of our Raman data, the multiple excitations appear with a period of about 630 cm$^{-1}$ up to the fourth order; and, since we have unveiled that the first- and second-order Raman scattering have a common origin, and the excitation observed in the first order at around 655 cm$^{-1}$  must be shifted by about 25 cm$^{-1}$ from the central laser line. This is consistent with our analysis presented in Fig. 3(a). The ratio between the two main peaks  in the periodically repeated three-peak structure changes progressively in the multi-order scattering, from the dominating higher-frequency feature in the second-order process to the opposite case in the fourth-order process, whereas the two features are approximately equal in the third-order process. This might be associated with a progressive decrease of the $A_g$ contribution at 650  cm$^{-1}$ in the higher-order spectra, as observed in the second-order process (see the inset of Fig. 1(b)), where the decreased $A_g$ contribution at around 1300 cm$^{-1}$ against the enhanced relative contribution of $B_{1g}$ symmetry at around 1165 and 1285 cm$^{-1}$ can be observed. From this follows that the anomalous Raman scattering is essentially promoted by the $B_{1g}$ mode at 510  cm$^{-1}$, which appears periodically at around 1165, 1795, and 2425 cm$^{-1}$ in the multiple-order spectra.

\section{Discussion and conclusions \label{sec:level3}}

We have investigated a detailed temperature dependence of the first and higher-order Raman scattering in the detwinned LaMnO$_3$ crystal and, based on the results of this study, we can conclude that the multi-order Raman scattering arises from the series of transitions, and that the triplet structure observed in the Raman spectra in the 900 -- 1400 cm$^{-1}$ range is the second replica in the series. This is in contrast to the conclusions of Saitoh {\it et al.} \cite{Saitoh}, where the three peaks observed in the 900 -- 1400 cm$^{-1}$ range at around 1008, 1170, and 1290 cm$^{-1}$ were previously interpreted as a result of dispersion of an independent orbital excitation, related to the superexchange interaction. We show that the triplet structure, being shifted from the zero frequency by about $\Omega\simeq$ 25  cm$^{-1}$ in the $x'$ polarisation, is repeated in the multi-order spectra with a period of $\omega_1\simeq$ 630 cm$^{-1}$. In this case, the three pronounced excitations (numerated by $i$=1,2, and 3) in  the n$th$-order  Raman spectra ($n$=1,2,3,4,..) can be represented as $\omega_i^{(n)}=\Omega+\omega_i+(n-1)\omega_1$. For the $x'$-polarised spectra, measured on our detwinned LaMnO$_3$ crystal (see Fig. 4), we determine $\omega_1\simeq$ 630 cm$^{-1}$, $\omega_2\simeq$ 510 cm$^{-1}$, and $\omega_3\approx$ 320 cm$^{-1}$. We note that in the valid multiphonon scenario \cite{Allen,Grueniger}, a sum of the $k$ vectors of the phonons involved in the Raman scattering at $\omega_{i}^{(n)}$ must be $\approx0$. To satisfy these conditions, all the phonons, $\Omega$, $\omega_1$, $\omega_2$, and $\omega_3$, must be the zone-center phonons. However, we cannot find the $\Gamma$-point Raman and/or infrared mode at $\omega_1\simeq$ 630 cm$^{-1}$ \cite{note}. Thus, we assume that the observed effect may result from combinations of other phonons from the Brillouin zone, provided their total $k$ vector $\approx0$. It has been suggested  by Iliev   {\it et al.} that the feature at around  650 cm$^{-1}$, replicated in the multi-order Raman scattering, corresponds to the maximum of the phonon density-of-states (PDOS) of oxygen vibrations, which is in good agreement with the lattice-dynamics theoretical calculations \cite{IlievPDOS,IlievPDOS1}. According to the results of our Raman study  on the detwinned LaMnO$_3$ crystal in the multiphonon spectral range, we propose that the PDOS feature has a period of $\omega_1\simeq$ 630 cm$^{-1}$ in the $x'$ polarisation, and that the strong peak at $\omega_2\simeq$ 510 cm$^{-1}$ is dominated by  the contribution from the zone-boundary $B_{1g}$ phonons to the PDOS.

We still need to provide an explanation for the shift of the multi-order Raman scattering from the zero frequency by $\Omega\simeq$ 25 cm$^{-1}$, observed in this study (see Fig. 4 and the inset). We suggest that this shift could result from a low-energy mode, allowed by symmetry; in this case, the periodically repeated oxygen PDOS feature with a period of $\omega_1\simeq$ 630 cm$^{-1}$ could be sidebands, activated by some process, and not allowed by symmetry. If this is true, one should be able to observe the low-energy mode at $\Omega\simeq$ 25 cm$^{-1}$ in a far-infrared experiment. We are aware of the experimental evidence of the mode at about 25 cm$^{-1}$ in doped rare-earth manganites. In Pr$_{0.3}$Ca$_{0.7}$MnO$_3$ the observed low-energy absorption peak at 16 -- 24 cm$^{-1}$ was assigned to a collective excitation of the charge-density-wave state \cite{Kida}.  This hypothesis, however, was challenged in a recent study of the low-energy excitations in La$_{0.25}$Ca$_{0.75}$MnO$_3$, where a resonance-like absorption observed at about 25 cm$^{-1}$ was associated with the optically activated acoustic phonon due to a fourfold superstructure along the $y$ axis in the charge-ordered state below $T_{CO}\approx 240$ K \cite{Zhukova}. However, the low-energy mode observed in La$_{0.25}$Ca$_{0.75}$MnO$_3$ at around 25 cm$^{-1}$ appears at temperatures below $T_{\rm N} \approx 140$ K \cite{Zhukova} and seems to be not directly related to the formation of the charge-ordered state in the doped manganites.
A far-infrared-active broad relaxation-like mode, excited by the $x$-polarised electric a.c. component of the radiation at about the same frequency in multiferroic manganites  TbMnO$_3$ (at 20 $\pm$ 3 cm$^{-1}$) and GdMnO$_3$ (at 23 $\pm$ 3 cm$^{-1}$), was ascribed to an electro-active magnon excitaton \cite{Pimenov}. The mode intensity increases with decreasing temperature in the low-temperature magnetic phase, where it can be suppressed by an external $z$-axis polarised magnetic field of an order of several Tesla \cite{Pimenov}. In multiferroics of more complex mixed-valence manganites of $R$Mn$_2$O$_5$ ($R$=Pr,Sm,Gd,Tb) a strong featureless band around these frequencies in the infrared reflectivity spectra was associated with collective electronic excitations of partially delocalised $e_g$ electronic orbitals, participating in dynamical multiferroic coupling in the Jahn-Teller distortion of the pyramid dimer \cite{Massa}. The transmission measurements detected this mode in TbMn$_2$O$_5$ and YMn$_2$O$_5$ at 9.6 cm$^{-1}$ and 7.2 cm$^{-1}$, respectively \cite{Sushkov}. Additionally, the high-frequency mode, corresponding to the Raman Mn-O stretching mode, appears in the infrared spectra at the ferroelectric transition in TbMn$_2$O$_5$ \cite{Drew}. These experimental evidences certify that the low-energy mode gains its spectral weight and becomes activated in doped rare-earth manganites and Mn-based multiferoics. The anomalous Raman scattering was also observed in the mixed $B$-site orthorhombic perovskites of LaFe$_{1-x}$Cr$_{x}$O$_3$, activated by the photon-induced electronic charge transfer between the neighbouring transition-metal ions of Cr$^{3+}$ (3$d^3$) and Fe$^{3+}$ (3$d^5$) \cite{Andreasson1,Andreasson2}. The observed anomalous Raman scattering was shown to be critically dependent on the presence of both $B$-site type cations, as well as the doping level, in this system with strong electron-phonon interaction, locally disturbing coherence in the electronic and vibronic structure \cite{Andreasson1,Andreasson2}. In a view of the results obtained in the present study, it will be interesting to look for the infrared-active low-energy mode in the mixed $B$-site orthorhombic perovskites of LaFe$_{1-x}$Cr$_{x}$O$_3$.

We argue that the low-energy mode at $\Omega\simeq$ 25  cm$^{-1}$ could be a tunneling transition in LaMnO$_3$, and the periodically repeated oxygen PDOS feature with a period of $\omega_1\simeq$ 630 cm$^{-1}$ could be the sidebands, activated by the electron tunneling between the potential energy minima generated near the JT Mn$^{3+}$ ion by the lattice anharmonicity. Indeed, first of all, we note that the frequency of the mode at around 25 cm$^{-1}$ is consistent with the tunneling frequencies in the dynamic limit of the Jahn-Teller effect in, for example, a $^2$D system of 3 $d^9$ MgO : Cu$^{2+}$ and 3 $d^1$ CaF$_2$ : Sc$^{2+}$ \cite{HayesStoneham}. Secondly, in our recent optical study of LaMnO$_3$ we have established that the kinetic energy changes of the $e_g$ electrons from the low- to high-temperature limit, associated with the high-spin $d^4d^4 \Longleftrightarrow d^3d^5$ electronic transition, seen in the optical spectra at 2 eV, is about 55 meV \cite{KovalevaLMO_prb}. This energy is comparable with the orbital ordering temperature, $T_{OO}$ = 780 K = 65 meV, and much higher than the energy of the average lattice anharmonicity barriers between the  potential energy minima of about 100 cm$^{-1}$ (12 meV). Therefore, the superexchange interaction  may be crucial in driving the $e_g$ electrons to the regime of the dynamic JT effect, with affordable barriers for the tunneling between the potential energy minima.

Finally we note that the dynamic regime of the Jahn--Teller effect could result in the anomalous multi-order Raman scattering in manganites. Indeed, we have discovered in the present study that aside from the oxygen-phonon density-of-states feature, there is an additional component to the multi-order scattering, which exhibits a different behavior. This character may arise from an admixture of the electronic component in the case of mixing between the electronic and vibrational states. For a weak vibronic interaction, when the barriers between the minima are less than the characteristic vibrational energy, a special coupling between low-frequency electronic motions and vibrational modes takes place \cite{Moffitt,Bersuker}. In the linear approximation, the eigenvalues of the system are those of a doubly-degenerate harmonic oscillator, $E_n=(n+1)\hbar\nu$, where $n$ is a positive integer and the $nth$ level is ($n$+1)-fold degenerate.  In the second order of the perturbation theory, the degeneracy can be removed and all levels, except the ground state level with $n$=0, become split. This splitting is similar to the effect of spin-orbit interaction in diatomic molecules, where for each $n$ there is the degeneracy with $-n, -n+2, ...,n$.

Further experiments are needed to investigate and verify the origin of the low-energy mode at about 25 cm$^{-1}$ and the low-energy electron-vibronic coupling in manganites, which is indicative of the dynamic regime of the Jahn--Teller effect in manganites.

\ack
The authors acknowledge helpful discussions with M Cardona, K P Meletov, and \mbox{I I Tartakovskii}.

This work was supported by the ESF network-program AQDJJ, the Royal Society of London (grant JP 090710), the Russian Foundation for Basic Research (projects10-02-92600-KO and 11-02-00708), Program of the Presidium RAS ``Quantum mesoscopic and disordered systems'', grant
TA01010517 of the TACR (Technology Agency of the Czech Republic), and AV0Z10100522 of the ASCR (Academy of Sciences of the Czech Republic).

\newpage

\begin{figure}[tbp]
\begin{center}
\includegraphics*[width=120mm]{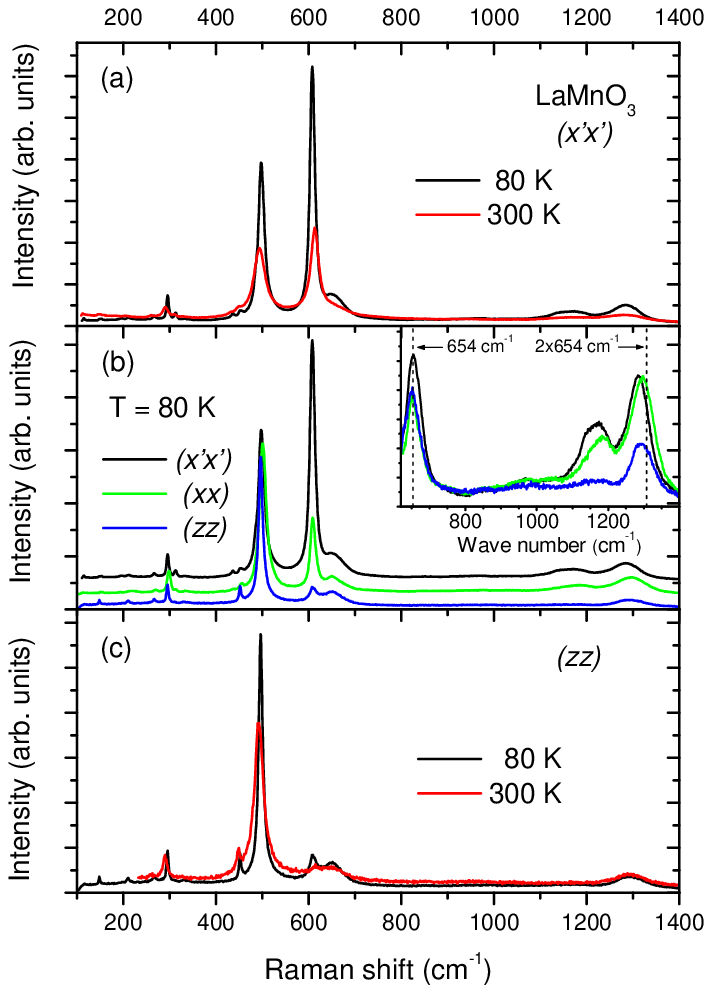}
\end{center}
\caption{(Colour online) The low-temperature (80 K) and room-temperature (300~K) Raman spectra, measured on the oriented $xz$ surface of the detwinned LaMnO$_3$ crystal, with the incident laser light polarised along (a) the $x'$ direction rotated by 45$^\circ$ with respect to the main axes, (c) the $z$ axis. (b) The low-temperature (80 K) Raman spectra measured in the $x,x'$ and $z$ polarisations and normalised to the $A_g$ (495 cm$^{-1}$) mode are shifted along the vertical axis. Inset: The zoom-in of the polarised first-order Raman spectra (separated from the symmetry-allowed lattice phonons) and the second-order Raman spectra.}
\label{Fig1}
\end{figure}

\begin{figure}[tbp]
\begin{center}
\includegraphics*[width=130mm]{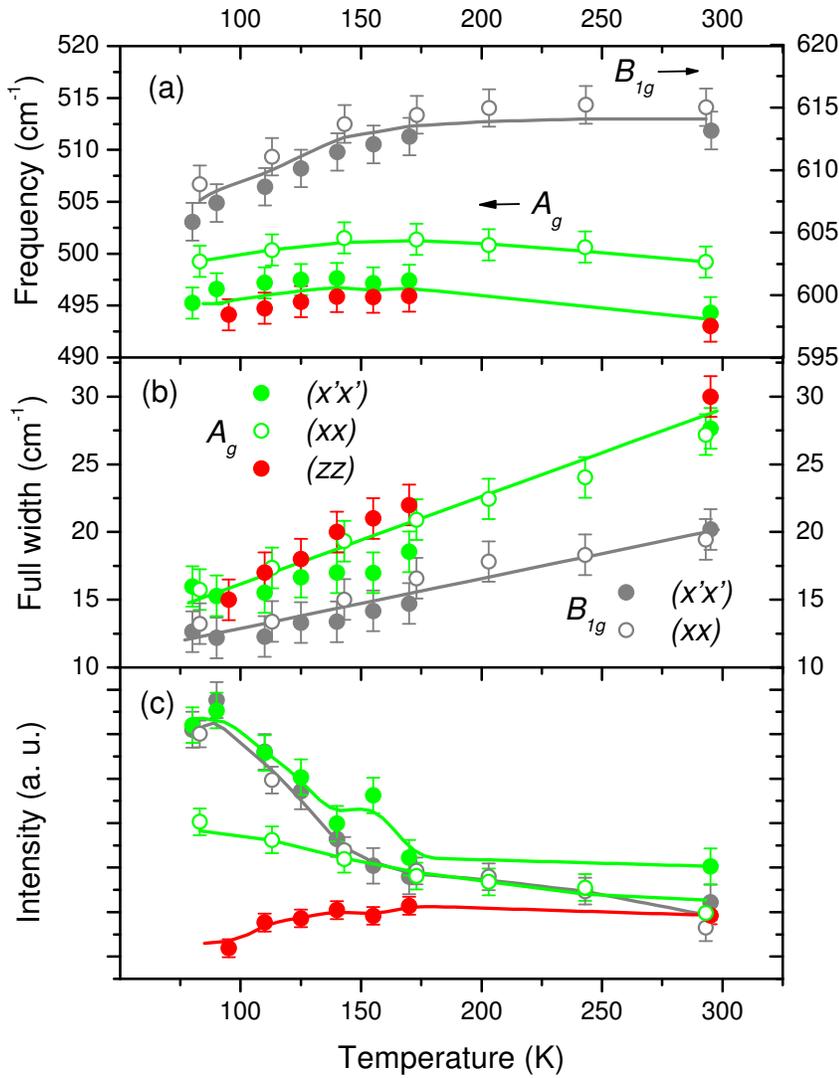}
\end{center}
\caption{(Colour online) Temperature and polarisation dependences of (a) the resonant frequencies, (b) the full widths at half maximum (FWHM), and (c) the scaled integrated intensities of the JT $A_g$ and $B_{1g}$ modes, resulting from the fit of the Raman spectra with Lorentzian bands. Solid curves in panels (a,c) are the guidelines for the dependencies averaged within the error bars. Temperature dependences of the FWHM of the JT $A_g$ and $B_{1g}$ modes in (b) are fitted by the linear functions.}
\label{Fig2}
\end{figure}

\begin{figure}[tbp]
\begin{center}
\includegraphics*[width=125mm]{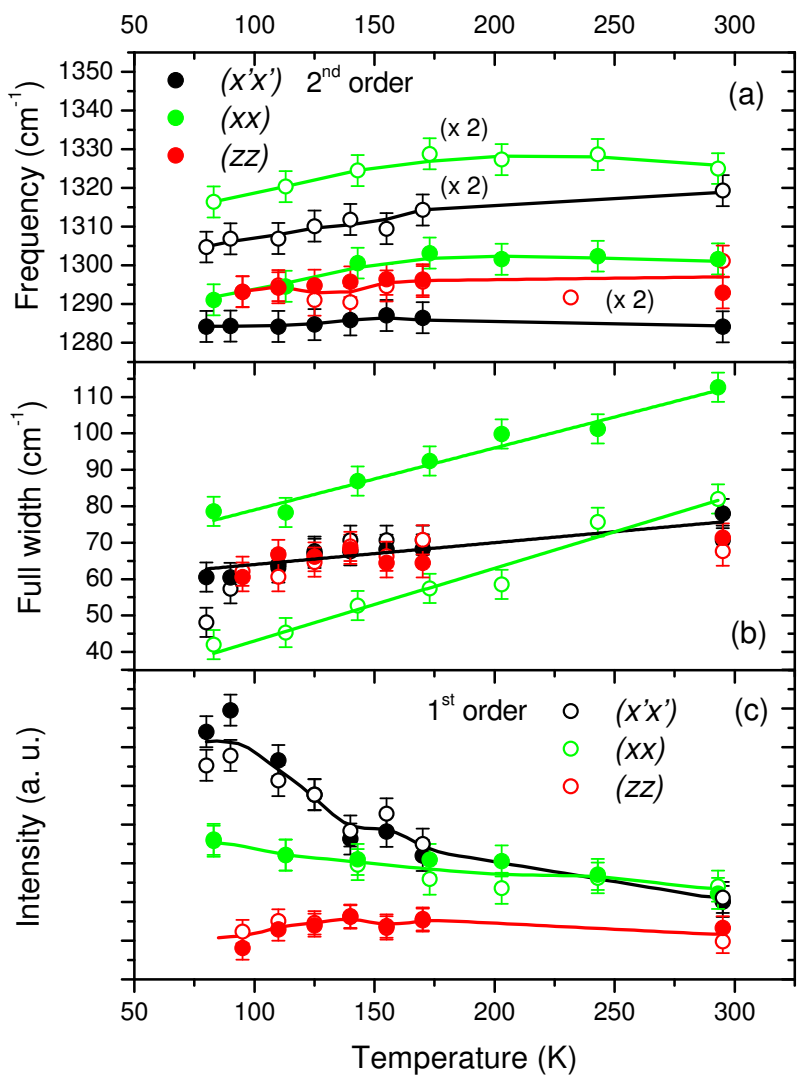}
\end{center}
\caption{(Colour online) Temperature and polarisation dependences of the first-order peak at about 650 cm$^{-1}$ and the second-order peak at about 1290 cm$^{-1}$, resulting from the fit of the Raman spectra with Lorentzian bands: (a) the resonant frequencies of the first-order peak multiplied by two are compared with those of the second-order peak, (b) full widths at half maximum (FWHM) are fitted by the linear temperature dependences, and (c) scaled integrated intensities. Solid curves in panels (a,c) are the guidelines for the dependences averaged within the error bars.}
\label{Fig3}
\end{figure}

\begin{figure}[tbp]
\begin{center}
\includegraphics*[width=115mm]{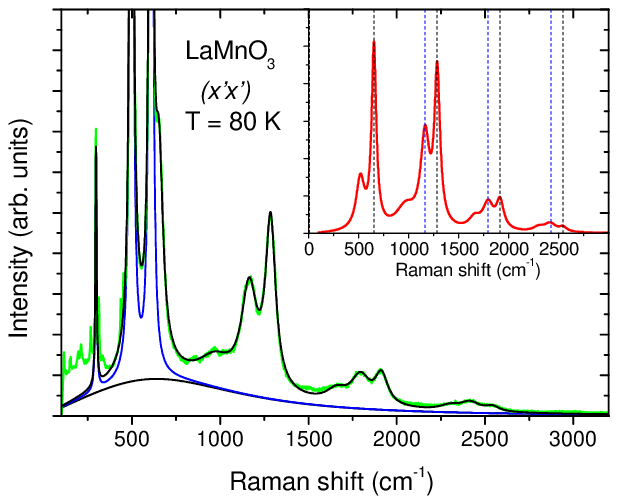}
\end{center}
\caption{(Colour online) The $x'$-polarised Raman spectra, measured in the multiphonon range on the detwinned LaMnO$_3$ crystal at T = 80 K (green curve). The solid curve, superimposed with the Raman spectra, demonstrates the accuracy of the fitting with the Lorentzian bands. Inset: the result of the fitting, subtracted from the wide background and the first-order Raman lattice phonons. The multiple excitations appear with a period of 630 cm$^{-1}$ up to the fourth order, being shifted from the zero frequency by about 25 cm$^{-1}$.}
\label{Fig4}
\end{figure}


\section*{References}
\begin{thebibliography}{99}


\bibitem{Martin-Carron} Mart\'in-Carr\'on L and de Andr\'es A 2004 {\it Phys. Rev. Lett.} {\bf 92} 175501

\bibitem{Jandl} Jandl S, Laverdi\`ere J, Mukhin A A, Ivanov V Yu and Balbashov A M 2006 {\it Physica} B \mbox{{\bf 381} 214}

\bibitem{Choi} Choi K-Y, Lemmens P, Sahaouri T, G\"untherodt G, Pashkevich Yu G, Gnezdilov V P, \mbox{Reutler P}, Pinsard-Gaudart L, B\"uchner L B and Revcolevschi A 2005 {\it Phys. Rev.} B {\bf 71} 174402

\bibitem{Choi1} Choi K-Y, Lemmens P, G\"untherpdt G, Pashkevich Yu G, Gnezdilov V P, Reutler P, Pinsard-Gaudart L, B\"uchner B and A. Revcolevschi 2005 {\it Phys. Rev.} B {\bf 72} 024301

\bibitem{Laverdiere} Laverdri\`ere J, Jandl S, Mukhin A A and Ivanov V Yu 2006 {\it Eur. Phys. J.} B {\bf 54} 67

\bibitem{SarfattStoneham} Sarfatt J and Stoneham A M 1967 {\it Proc. Phys. Soc.} {\bf 91} 214

\bibitem{Allen_nature} Allen P B and Perebeinos V 2001 {\it Nature} {\bf 410} 155

\bibitem{Saitoh} Saitoh E, Okamoto S, Takahashi K T, Tobe K, Yamamoto K,
Kimura T, Ishihara S, \mbox{Maekawa S} and Tokura Y 2001 {\it Nature} {\bf 41} 180

\bibitem{Goodenough} Goodenough J B 1955 {\it Phys. Rev.} {\bf 100}
564

\bibitem{Kanamori} Kanamori J 1960 {\it J. Appl. Phys.} {\bf 31} S14

\bibitem{KugelKhomskii} Kugel K I and Khomskii D I 1982 {\it Sov. Phys. Uspekhi} 1982 {\bf 25} 231 [1982 {\it Usp. Fiz. Nauk} \mbox{{\bf 136} 621]}

\bibitem{Oles} Ole\'s Andrzej M, Khaliullin G, Horsch P and Feiner L F 2005 {\it Phys. Rev.} B {\bf 72} 214431

\bibitem{Kaplan} Kaplan M D and Vekhter B G  1995 {\it Cooperative Phenomena in Jahn-Teller Crystals} (New York: Plenum Press)

\bibitem{Maekawa} Okamoto S, Ishihara S, and Maekawa S 2002 {\it Phys. Rev.} B {\bf 65} 144403

\bibitem{Okamoto} Okamoto S, Ishihara S and Maekawa S 2002 {\it Phys. Rev.} B {\bf 66} 014435

\bibitem{Brink} Van den Brink J 2001 {\it Phys. Rev. Lett.} {\bf 87} 217202

\bibitem{Allen} Allen P B and Perebeinos V 1999 {\it Phys. Rev. Lett.} {\bf 83} 4828

\bibitem{Allen1} Perebeinos V and Allen P B 2001 {\it Phys. Rev.} B {\bf 64} 085118

\bibitem{kovaleva_jetp} Kovaleva N N, Gavartin J L, Shluger A L,
Boris A V and Stoneham A M 2002 {\it JETP} {\bf 94} 178 [2002 {\it Zh. Eksp. Teor. Fiz.} {\bf 121} 210]

\bibitem{KovalevaLMO_prl} Kovaleva N N, Boris A V, Bernhard C, Kulakov A, Pimenov A, Balbashov A M, \mbox{Khaliullin G} and Keimer B 2004 {\it Phys. Rev. Lett.} {\bf 93} 147204

\bibitem{KovalevaLMO_prb} Kovaleva N N, Ole\'s Andrzej M, Balbashov A M,
Maljuk A, Argyriou D N, Khaliullin G and Keimer B 2010 {\it Phys. Rev.} B {\bf 81} 235130

\bibitem{IlievPDOS} Iliev M N, Abrashev M V, Popov V N and Hadjiev V G 2003 {\it Phys. Rev.} B {\bf 67} 212301

\bibitem{IlievPDOS1} Iliev M N, Hadjiev V G, Litvinchuk A P, Yen F, Wang  Y-Q, Sun Y Y, Jandl S, Laverdi\`ere J, Popov V N and Gospodinov M M 2007 {\it Phys. Rev.} B {\bf 75} 064303

\bibitem{Goldstone} Goldstone J, Salam A and Weinberg S 1962 {\it Phys. Rev.} {\bf 127} 965

\bibitem{Stoneham} Stoneham A M 1975 {\it Theory of Defects in Solids} (Oxford: Clarendon Press) p 189

\bibitem{HayesStoneham} Hayes W and Stoneham A M 1985 {\it Defects and Defect Processes in Nonmetallic Solids} (New York, Chichester, Brisbane, Toronto, Singapore: John Wiley Sons) p 194

\bibitem{Bersuker} Bersuker I B, Vekhter B G and Ogurtsov I I 1975 Sov. Phys. Uspekhi {\bf 18} 569 [1975 {\it Usp. Fiz. Nauk} {\bf 116} 605]

\bibitem{FeinerOles} Feiner L F and Ole\'s A M 1999 {\it Phys. Rev.} B {\bf 59} 3295

\bibitem{Paolasini} Paolasini L, Caciuffo R, Sollier A, Ghigna P and Altarelli M 2002 {\it Phys. Rev. Lett.} {\bf 88} 106403

\bibitem{Deisenhofer} Deisenhofer J, Leonov I, Eremin M V, Kant Ch, Ghigna P, Mayr F, Iglamov V V, \mbox{Anisimov V I} and van der Marel D 2008 {\it Phys. Rev. Lett.} {\bf 101} 157406

\bibitem{Abbamonte} Lee J C T, Yuan S, Lal S, Joe Y, Gan Y, Smadici S, Finkelstein K, Feng Y, Rusydi A, \mbox{Goldbart P M}, Cooper S L and Abbamonte P 2012 {\it Nature Physics} {\bf 8} 63

\bibitem{Murakami} Murakami  Y, Hill J P, Gibbs D, Blume M, Koyama I, Tanaka M, Kawata H, Arima T, \mbox{Tokura Y}, Hirota K   and Endoch Y 1998 {\it Phys. Rev. Lett.} {\bf 81} 582

\bibitem{Granado} Granado E, Moreno N O, Garc\'ia A, Sanjurjo J A, Rettori C, Torriani I, Oseroff S B, \mbox{Neumeier J J}, McClellan K J, Cheong S-W and Tokura Y 1998 {\it Phys. Rev.} B {\bf 58} 11435

\bibitem{Granado1} Granado E, Garcia A, Sanjurjo J A, Rettori C, Torriani I, Prado F, Sanchez R D, Canairo A and Oseroff S B 1999 {\it Phys. Rev.} B {\bf 60} 11879

\bibitem{Kida} Kida N and Tonouchi M 2002 {\it Phys. Rev.} B {\bf 66} 024401

\bibitem{Zhukova} Zhukova E, Gorshunov B, Zhang T, Wu Dan, Prokhorov A S, Torgashev V I, Maksimov E G and Dressel M 2010 {\it EPL} {\bf 90} 17005

\bibitem{Moffitt} Moffitt W and Liehr A D 1956 {\it Phys. Rev.} {\bf 106} 1195

\bibitem{Balbashov} Balbashov A M, Karabashev S G, Mukovsky Ya M and Zverkov S A 1996 {\it J. Cryst. Growth} {\bf167} 365

\bibitem{Smirnova} Smirnova I S 1999 {\it Physica} B {\bf 262} 247

\bibitem{Iliev} Iliev M N, Abrashev M V, Lee H-G, Popov V N, Sun Y Y, Thomsen C, Meng R L and \mbox{Chu C W} 1998 {\it Phys. Rev.} B {\bf 57} 2872

\bibitem{KovalevaYTO} Kovaleva N N, Boris A V, Capogna L, Gavartin J L,
Popovich P, Yordanov P, Maljuk A, Stoneham A M and B. Keimer 2008 {\it Phys. Rev.} B {\bf 79} 045114

\bibitem{Krueger} Kr\"uger R, Schulz B, Naler S, Rauer R, Budelmann D, B\"ackstr\"om J, Kim K H, Cheong S-W, Perebeinos V and R\"ubhausen M 2004 {\it Phys. Rev. Lett.} {\bf 92} 097203

\bibitem{Tobe} K. Tobe, Kimura T, Okimoto Y and Tokura Y 2001 {\it Phys. Rev.} B {\bf 64} 184421

\bibitem{Quijada} Quijada M A, Simpson J R, Vasiliu-Doloc L, Lynn J W, Drew H D, Mukovskii Y M and Karabashev S G 2001 {\it Phys. Rev.} B {\bf 64} 224426

\bibitem{Grueniger} Gr\"uninger M, R\"uckamp R, Windt M, Reutler P, Zobel C, Lorentz T, Freimuth A and Revcolevschi A 2001 {\it Nature} {\bf 418} 39

\bibitem{note} Our ellipsometry measurements on the same (detwinned)
LaMnO$_3$ single crystal have shown that the excitation at $\sim$ 650 cm$^{-1}$ is also in-plane polarised

\bibitem{Tofield} Tofield B C and Scott W R 1974 {\it J. Solid State Chem.} {\bf 10} 183

\bibitem{Roosmalen} Van Roosmalen J A M, Cordfunke E H P, Helmholdt R B and Zandbergen H W 1994 {\it J. Solid State Chem.} {\bf 110} 100

\bibitem{Pimenov} Pimenov A, Mukhin A A, Ivanov V Yu, Travkin V D, Balbashov A M, Loidl A 2006 {\it Nature Physics} {\bf 2} 97

\bibitem{Massa} Massa N E, Garc\'ia-Flores A F, Meneses D S, Campo L, Echegut P, Fabbris G F L, Mart\'inez-Lope M J, Alonso J A 2012 {\it J. Phys.: Condens. Matter} {\bf 24} 195901

\bibitem{Sushkov} Sushkov A B, Aguilar R V, Park S, Cheong S -W and Drew H D 2007 {\it Phys. Rev. B} {\bf 98} 027202

\bibitem{Drew} Aguilar R V, Sushkov A B, Park S, Cheong S -W, Drew H D 2006 {\it Phys. Rev. B} {\bf 74} 184404

\bibitem{Andreasson1} Andreasson J, Holmlund J, Rauer R, K\"all M, B\"orjesson L, Knee C S, Eriksson A K, Eriksson S-G, R\"ubhausen M, Chaudhury R P{\it Phys. Rev. B} {\bf 78} 235103

\bibitem{Andreasson2} Andreasson J, Holmlund J, Singer S G, Knee C S, Rauer R, Schulz B, K\"all M, R\"ubhausen M,  Eriksson S-G, B\"orjesson L, Lichtenstein A 2009 {\it Phys. Rev. B} {\bf 80} 075103



\end{thebibliography}
\end{document}